**Title: Authoritarian Governments Appear to Manipulate COVID Data**

**Authors:**

Mudit Kapoor[1*]

Anup Malani[2,*]

Shamika Ravi[3]

Arnav Agarwal[4]

**Affiliations:**

[1] Indian Statistical Institute, New Delhi, India

[2] University of Chicago, Chicago, US

[3] Independent economist, New Delhi, US

[4] Harvard University, Boston, US

* Corresponding authors

**Author contributions:**

M.K. and A.M. contributed to conceptualization, methodology, validation, formal analysis, writing, and visualization.

S.R. contributed to conceptualization and methodology.

A.A. contributed to investigation and data curation.

**Competing interests:**

The authors declare no competing interests. This project received no funding.

**Materials & correspondence:**

Correspondence and data requests should be addressed to Anup Malani.

**Abstract:**

Because SARS-Cov-2 (COVID-19) statistics affect economic policies and political outcomes, governments have an incentive to control them. Manipulation may be less likely in democracies, which have checks to ensure transparency. We show that data on disease burden bear indicia of data modification by authoritarian governments relative to democratic governments. First, data on COVID-19 cases and deaths from authoritarian governments show significantly less variation from a 7 day moving average. Because governments have no reason to add noise to data, lower deviation is evidence that data may be massaged. Second, data on COVID-19 deaths from authoritarian governments do not follow Benford's law, which describes the distribution of leading digits of numbers. Deviations from this law are used to test for accounting fraud. Smoothing and adjustments to COVID-19 data may indicate other alterations to these data and a need to account for such alterations when tracking the disease.

**Main text:**

The World Health Organization declared SARS-CoV-2 (COVID-19) a pandemic on March 11, 2020. As of June 30, there have been more than ten million cases and in excess of five hundred thousand COVID-19 deaths[1]. One feature that makes COVID-19 unique among recent epidemics is the excessive burden of the disease in democratic countries. Using the Economist Intelligence Unit's (EIU) 2019 Democracy Index[2], which classifies governments into four regimes (authoritarian, hybrid, flawed democracy, and full democracy), we find that although democracies account for 48% of the world's population, they account for approximately 75% and 89% of total cases and deaths, respectively. Figure 1 compares COVID-19 burden per million people across regimes through June 30. For authoritarian regimes the median value of cumulative cases per million people was 324 (Interquartile range [IQR]; 56-1529), whereas for full democracies it was 2896 (IQR; 1308-5222). Comparing cumulative deaths per million people; for authoritarian regimes the median value was 8.6 (IQR; 0.9-29.3), while for full democracies the median was 130.8 (IQR; 28.4-355.9). Moreover, the growth in cases and deaths is more pronounced in democratic regimes, in particular, full democracies, as compared to other regimes. Likewise, the case fatality rate (ratio of COVID-19 deaths to total cases) has risen faster in full democracies. Indeed, the CFR has declined in authoritarian regimes since late March (Figure S1).

There are several possible explanations for the high burden in democracies. First, democracies are on an average richer (higher per capita income and health expenditure as a percentage of gross domestic product) than other regimes. They can afford more tests, resulting in higher case and death counts. Second, democracies are more open to travel and trade. This facilitates the spread of COVID-19 across borders. Third, democracies may, idiosyncratically, have a larger elderly population, which is more vulnerable to COVID-19. Fourth, most democratic countries are north of 40° latitude. Fifth, perhaps authoritarian regimes have greater control over their population. They may be better able to enforce social distancing and limit mobility, both of which reduce spread of the disease.

These explanations presume that the data on COVID-19 burden are reliable. However, the press has raised questions about the credibility of COVID-19 data reported by countries. Stories regarding data manipulation have emerged for China [3], Iran [4], Indonesia [5], and the US [6]. Therefore, it is important statistically to investigate the reliability of COVID-19 data that is being reported across regimes.

In democracies, with freedom of the press, separation of power, and an active opposition, there may exist checks and balances that prevent governments from manipulating the data. Authoritarian regimes have greater latitude to manipulate data. Such governments have been criticized, however, for manipulating other types of data [7-10]. These governments have an incentive to use information as a form of social control [11-15].

Here we show evidence of manipulation of COVID-19 data by authoritarian regimes relative to democratic regimes. First, data from authoritarian governments show significantly less variation from a 7 day moving average. Because governments have no reason to add noise to data, lower deviation is evidence that data may be massaged. Second, data from authoritarian governments do not follow Benford's law, which describes the distribution of leading digits of non-manipulated numbers. These discrepancies do not provide direct evidence that the lower burden on authoritarian governments is due to data manipulation. However, they do provide indirect evidence: these modifications likely have a purpose and a plausible reason is suppressing bad news.

Ensuring the credibility of data isn't a coronavirus specific concern. Data manipulation has been a perennial concern in public health and economics. There are notable instances of data fabrication in research [16], disease surveillance [17,18], and measurement of economic conditions [7-10].

There are many statistical methods for detecting fraud [16,19,20]. Here we focus on two types of tests. One compares moments of the distribution of data across sites [19,21-23], specifically variance [16,19,20,22,24-26]. The other looks at digit preference that deviates from Benford's law [19,20,27,28].

**Results**

**Insufficient variation around moving average (Test 1).** There is a strong positive association between fluctuations in the COVID-19 data reported by different countries and their "democratic-ness". Figure 2 plots the natural logarithm of the mean of the squared deviation of daily cases and deaths per million people, respectively, from the 7 day moving average against the EIU's overall democracy index score. Not only do authoritarian regimes report fewer cases and deaths, there seems to be more random variation in the data in more democratic nations.

Aggregated data on Covd-19 across all countries in each of 4 regime categories provides further visual evidence that there is less variation in case data in authoritarian regimes. Figures S2a & S2b plots daily cases and deaths per million people around a 7-day centered moving average for those indicators, respectively, for each regime type. In addition to a lower rate of reported cases and deaths, there is almost no fluctuation in the data from authoritarian or hybrid regimes. Variation in the data appears to increase as one moves to a higher category of democratic-ness.

Regression analysis (Table 1) suggest that each unit increase in the EIU Democracy score is associated with a 0.25 log point (95% Confidence Interval[CI]; 0.11-0.40) increase in the squared deviation of daily cases and 0.29 (95% CI; 0.19-0.39) increase in the squared deviation of daily deaths, respectively per million people. We obtain similar significant results when we use other democracy indices from Freedom House, Varieties of Democracy Index, and Polity5 measures of political regimes.

Although it is unlikely that features that affect the level of COVID-19 burden affect the variation in that burden, we estimate a version of the regression in Table 1 with controls for GDP per capita, health and trade as a percent of GDP, share of population over 65 and an indicator for countries above 40 degrees latitude. While greater democratic-ness is no longer associated with additional variability in cases, it continues to be associated with significantly greater variability in daily deaths per million people (Table S1).

**Compliance with Benford's law (Test 2)**. Figure 3 presents the results of our analysis for cumulative case and death data when our screening criteria is that growth in the 7 day centered moving average is greater than 7.5%. (Results for tests for other screening criteria are presented in Tables S2 are roughly consistent.) One cannot reject that Benford's Law describes the distribution of first digits for cumulative cases for all regime types for p value less than 1%. However, one can reject the Benford's law that describes the distribution of first digits for cumulative deaths at p value less than 1% for cumulative deaths for authoritarian regime, hybrid regimes, and flawed democracy, while it cannot be rejected for full democracies.

**Validation with ECDC data**. All of the analysis reported above were also conducted with data from the ECDC and the results are very similar.

**Discussion**

Analysis of compliance with Benford's law suggests data from authoritarian regimes, hybrid regimes, and flawed democracy on cases comply but not for deaths, while for full democracies the data complies with Benford's law, both for cases and deaths. Higher deaths may be more politically salient and, therefore, subject to manipulation. First, because the infection fatality rate of COVID-19 is close to 1%, cases are less consequential than COVID-19 deaths. Second, deaths better reflects state capacity than cases. Total cases are largely determined by transmissibility and infectiousness of the disease, and the total number of tests. Total deaths are influenced by, in addition to these factors, the health infrastructure, including availability of medical personnel and beds. Governments may be able credibly to blame low levels of testing on global shortages rather than government policy. Personnel and beds, however, require long term investments in medical education and construction. Therefore, a high death rate may imply the government has performed poorly for some time.

This study has several limitations. One is that, while we establish an association between data smoothening and government regimes, there may be potential confounders not included here that could alter the conclusions of the study. Second, no causal link has been established between government regimes and data smoothening. Third, the study does not present methods to obtain less biased estimates of cases. Comparison of multiple sources of information or indirect methods of measuring COVID-19, such as SARI cases or orders of caskets, are worth exploring. A fourth limitation is that the paper presents two major methods of detecting manipulation. There are others, and these may reveal a greater degree of manipulation.

The results here raise significant questions about the reliability of the data being reported by different countries and highlights the need for a degree of caution when making projections using such data. It may be appropriate to put in place systems for ongoing monitoring for fraud as are used for clinical trials [23,29-32].

**Materials and methods**

**Data.** Data on the type of regime in different countries come primarily from the Democracy Index 2019, by the Economist Intelligence Unit (EIU)[2]. The EIU's Democracy Index 2019 includes data from expert assessments, public-opinion surveys (like the World Values Survey), voter turnout and the balance between the executive and legislative branches of government. The index has been used extensively in the literature as a measure of the state of democracy, in works examining anything from health services accessibility [33] to prosocial behavior [34] to trade [35]. The Index provides countries a score from 0 to 10 based on ratings across 60 indicators across 5 topics (electoral processes and pluralism, civil liberties, the functioning of government, political participation, and political culture). Countries are classified as full democracies (scores > 8), flawed democracies (scores in (6, 8]), hybrid regimes (scores in (4,6]) and authoritarian regimes (scores ≤ 4). Given the arbitrary and discontinuous nature of the boundaries between these categories, we also directly use the numerical scores in our empirical analyses. We also employ data from other measures of democracy, such as Freedom House's Democracy, the Varieties of Democracy Index, and the Polity5 of the polity project; these are described in the supplement.

For validating results from the Democracy Index, we use three other sources. The first is Freedom in the World 2020 from Freedom House [36], a rating of political rights and civil liberties in 195 countries and 15 territories. Freedom in the World uses information from consultations, official records, and on-the-ground research conducted by external analysts, expert advisers and Freedom House staff. The second source is data from Varieties of Democracy Institute [37] at the Department of Political Science at the University of Gothenburg, Sweden. The V-Dem Index aggregates information on more than 250 measures to rate countries between 0 and 1. The third source is the modified polity score from the Polity5 Project [38] by the Center for Systemic Peace that codifies the characteristics of authority across countries in the world. It ranges from -10 (strongly autocratic) to +10 (strongly democratic).

Country-level data on COVID-19 cases and deaths, as well as country latitude, are from the 2019 Novel Coronavirus Data Repository at the Johns Hopkins University Center for Systems Science and Engineering[1]. JHU's COVID repository aggregates information on cases and deaths from official and civil sources like the WHO, Center for Diseases Control and Prevention (CDC), media reports, local health departments and DXY, an online community for Chinese physicians and healthcare professionals. In this evolving situation, JHU's data stream has emerged as one of the most reliable for academic research, modeling and policy decisions.

For validating results from JHU data, we use data on cases and deaths from the European Centre for Disease Prevention and Control (ECDC). The ECDC data are similar to JHU, except that they do not contain presumptive positive cases, defined as cases that have been confirmed by state or local labs, though not by national labs such as the CDC [39].

We do not employ World Health Organization (WHO) data on COVID cases and deaths because of a change in the reporting time for WHO numbers on March 18, 2020 that makes it difficult to compare WHO number before and after that date. Aggregate WHO data, on the one hand, and JHU and ECDC data, on the other, are very similar, with the exception of the period from February 12-16, 2020. We choose to use ECDC data rather than WHO data to validate results using JHU data because of errors found in the WHO data [39].

Country-level demographic and economic information (country–level per capita income, health and trade expenditure as a proportion of GDP, and the share of population over age 65) for the year 2017/2018 are drawn from the World Bank Open Database [40]. Missing values were substituted with regional averages.

We used data from the 165 independent states and two territories for which the EIU produced scores. This covered more than 99% of the world's population. COVID-19 data was only available for 161 countries, Hong Kong was classified as part of China, and there was no data for Comoros, Lesotho, North Korea, Tajikistan, and Turkmenistan. These countries accounted for more than 99% of the total cases and deaths across the world.

**Data availability**. All the data used for this study are publically available and will be posted, along with code for all statistical analyses, will be posted in a Github repository by the corresponding author.

**Insufficient variation around moving average (Test 1)**. One method of detecting data manipulation is to look for abnormal statistics (such as with the moments of the distribution) of the variable [19,21-23]. It is difficult to identify abnormal means because one may not observe actual cases separately from the numbers reported by countries. A challenge for identifying abnormal variation in data is that there is no obvious baseline for normal variation. However, because the virus may not care about regime type, a comparison of variation across regime types may highlight abnormalities. In general, differences in variation across regime types cannot a priori distinguish whether one type suppressed variation or another type added variation. However, it is unlikely that higher variation is associated with manipulation, because countries gain little from adding variation to their data [16,19,20,22,24-26]. By contrast, manipulating data can lead to reduced variation if care is not taken to reintroduce "normal" levels of variation [41]. Therefore, we investigate whether authoritarian governments manipulate data by testing whether their COVID-19 data is "smoothened" relative to democratic governments.

To determine if the difference in data variation between authoritarian and democratic regimes is statistically significant, we employ regression analysis. Our dependent variable is a measure of variation in burden. We compose this variable in three steps. First, we calculate a 7 day centered moving average in daily cases (deaths) for each day in each country. Second, we calculate the square of the deviation of the observed daily cases (deaths) around that moving average for each country. Third, we add one to the squared deviation and divide that by population (millions) and then take the natural logarithm. Our treatment variable is either the country's score on the EIU's Democracy Index, Freedom House's Democracy, the Varieties of Democracy Index, or the Polity5 of the polity project. Our regressions also include a constant. While our observations are at the country day level, we cluster standard errors at the country level to account for autocorrelation in COVID-19 burden.

**Compliance with Benford's law (Test 2)**. A second method of detecting data manipulation is to see if data follow patterns that are common in non-manipulated data. One such pattern is that the leading significant digits of a number (or mantissa) has a distribution such that $\Pr(\text{mantissa} < t/10) = \log_{10} t$ for t in [1,10] [28]. Also known as Benford's Law, a wide assortment of data obey this law [42-45]. Data have been checked against this distribution to test for fraud in accounting data [46], campaign contributions [47] and scientific data [48]. We investigate whether governments manipulate data by testing whether the COVID-19 data on cumulative cases and deaths across different regimes (authoritarian, hybrid, flawed democracy, and full democracy) confirms to Benford's law.

Before we can test COVID-19 case and death data against Benford's Law, we must decide whether these data are appropriate to test against the law. A concern is that early during an epidemic and after infections plateau, the data will have a number of repeated numbers. These repeats may be the result of true case counts but still violate the law. Therefore, we look at portions of the time series of COVID-19 data during which cases and deaths are rising. Specifically, we test data ("screened data") during which the growth rate of the 7 day moving average of cases and deaths is greater than some cutoff k, where k is 5%, 7.5%, and 10%.

To implement the test, we look only at the first digit of the screened case and death data. According to Benford's law, Pr(first significant digit = d) = $\log_{10}(1+d^{-1})$, for d = 1,2,...,9. We group countries into the 4 regimes (authoritarian, hybrid, flawed democracy, and full democracy) defined by the EIU's democracy index. Within each category, we compare the observed frequency of each digit d in the case data against the frequency predicted by Benford's Law using a Pearson's chi-squared test.

**Tables and figures**

**Figure 1**

Boxplot of total cases and deaths per million people of countries across 4 types of regime (authoritarian, hybrid, flawed democracy and full democracy).

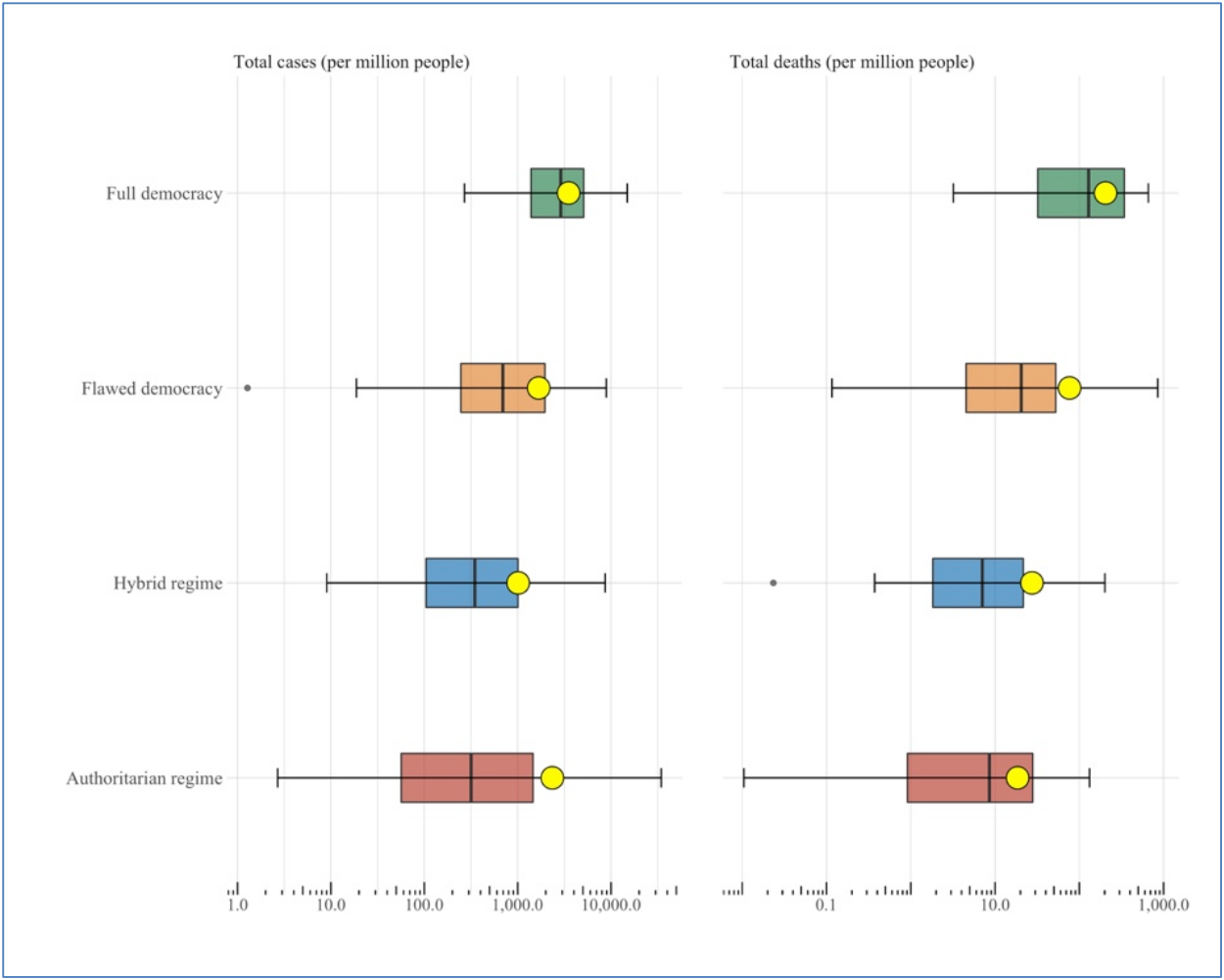

Notes. Regime classification is based on the Economist Intelligence Unit's 2019 Democracy Index. Case and death data are from Johns Hopkins University. Population data are from the World Bank Open Database. Upper (lower) hinge of the box is the $75^{th}$ ($25^{th}$) percentile. The black vertical line is the median and the yellow dot is the mean value. The black dots are the outliers.

**Figure 2**

Natural logarithm of the Mean of squared deviations of observed daily cases and deaths per million people from a 7-day centered moving average, by EIU democracy index score.

Daily cases

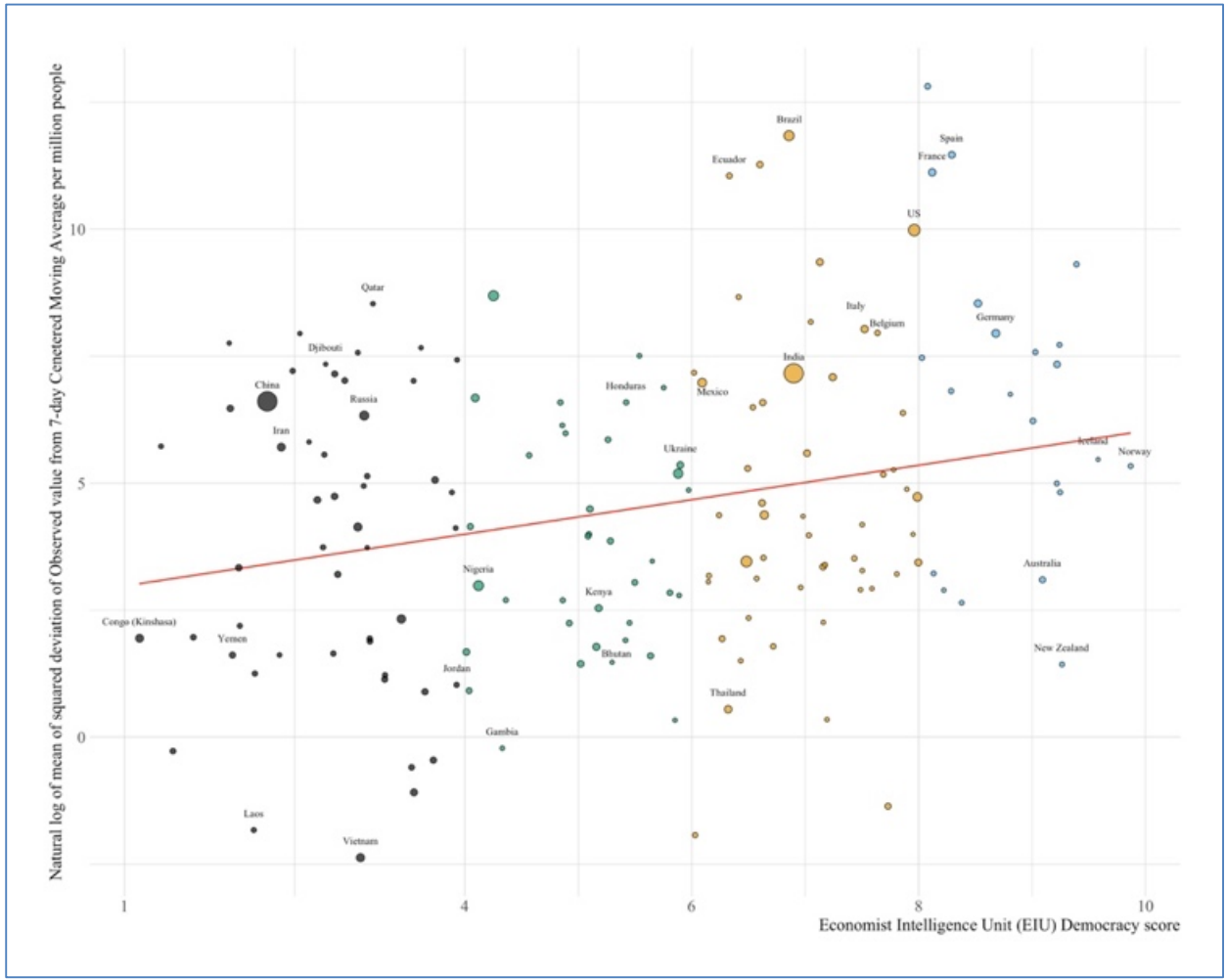

Daily deaths

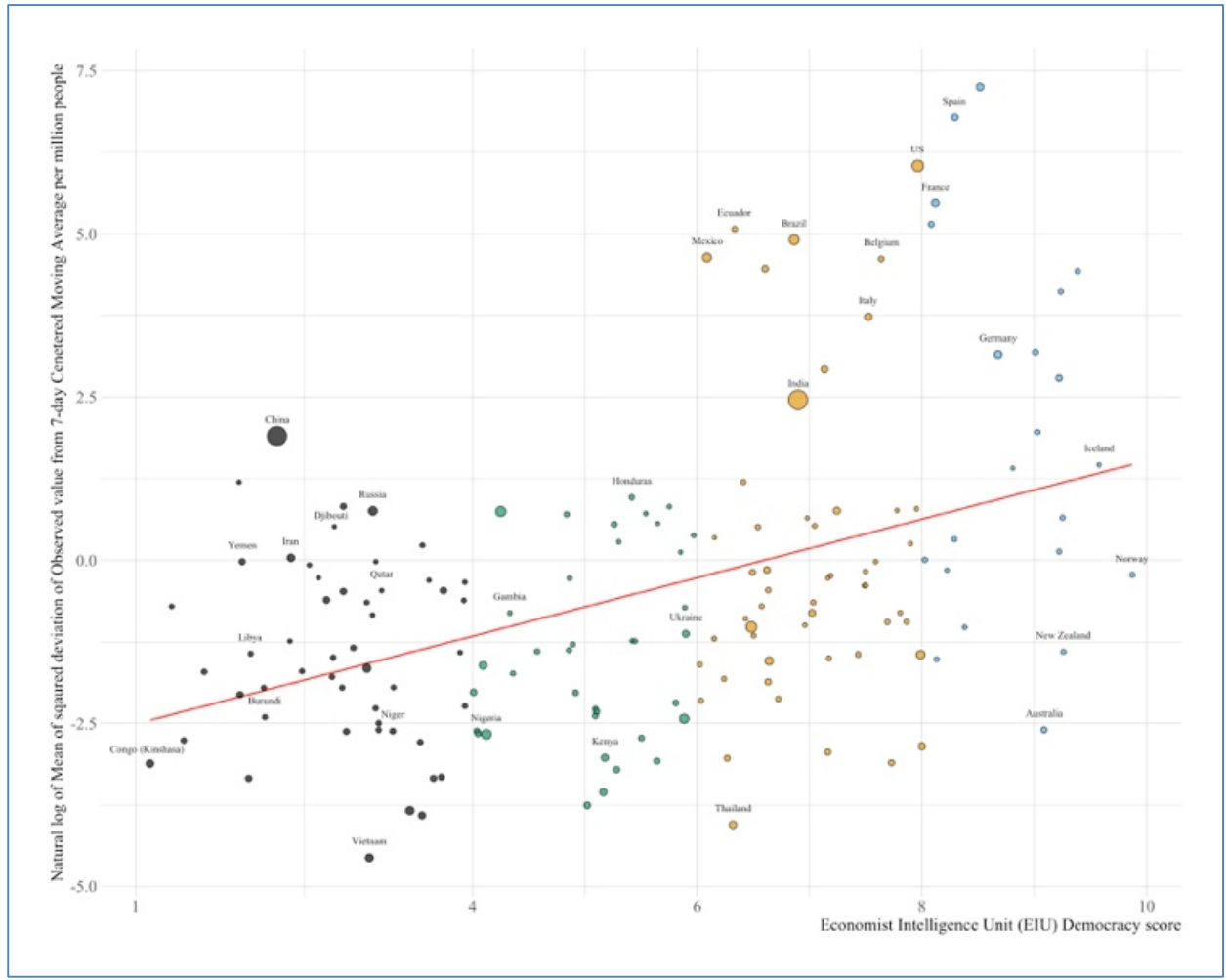

Notes. Case and deaths data are from Johns Hopkins University. The democracy index score is from the EIU's Democracy Index. We compute the 7 day centered moving average of daily cases and deaths. We compute the square of daily deviations of the observed cases (deaths) from the 7 day centered moving average and add one to it. Then for each country we divide this daily deviation by population per million, compute the mean for each country, and take the natural logarithm.

## Table 1

Ordinary least squares regression of deviations from a moving average on measures of democracy.

| | Daily cases | | | |
|---|---|---|---|---|
| Measure of democracy | EIU democracy score | FH democracy score | VD democracy score | Polity5 |
| Democracy | 0.25*** | 0.20*** | 0.23*** | 0.12 |
| | (0.11 - 0.40) | (0.08 - 0.32) | (0.10 - 0.37) | (-0.03 - 0.26) |
| Constant | 0.39 | 0.66 | 0.83** | 0.96 |
| | (-0.54 - 1.32) | (-0.15 - 1.48) | (0.14 - 1.52) | (-0.19 - 2.11) |
| Observations | 19,430 | 19,223 | 19,230 | 18,740 |
| R-squared | 0.02 | 0.03 | 0.03 | 0.01 |
| | Daily deaths | | | |
| Measure of democracy | EIU democracy score | FH democracy score | VD democracy score | Polity5 |
| Democracy | 0.29*** | 0.24*** | 0.28*** | 0.18*** |
| | (0.19 - 0.39) | (0.16 - 0.31) | (0.19 - 0.36) | (0.10 - 0.25) |
| Constant | -3.09*** | -2.84*** | -2.64*** | -2.80*** |
| | (-3.64 - -2.54) | (-3.31 - -2.36) | (-3.04 - -2.25) | (-3.39 - -2.22) |
| Observations | 19,430 | 19,223 | 19,230 | 18,740 |
| R-squared | 0.09 | 0.10 | 0.11 | 0.06 |

Note: *** $p<0.01$, ** $p<0.05$, * $p<0.1$. 95% Confidence intervals are in parenthesis. The errors are clustered at the country level. Our unit of analysis is the "country-date". The dependent variable is the natural logarithm of the squared deviation of the observed value from the 7 day centered moving average plus one per million people for each country on a daily basis, from the date when the first case was noted till June 30, 2020. Freedom House democracy score ranges from 0 to 100, to make it comparable to the EIU democracy score, the score is divided by 10. The VDEM score ranges from 0 to 1, to make it comparable to EIU democracy score, it is multiplied by 10. Similarly the modified polity5 score ranges from -10 (strongly autocratic) to +10 (strongly democratic), therefore, to make it comparable we add 10 to the score and divide it by 2.

**Figure 3**

Actual frequency of first significant digit in COVID-19 total cases and deaths during periods that 7 day centered moving average grows faster than 7.5% daily, frequency predicted by Benford's law, and test of the difference, by regime type.

Total cases

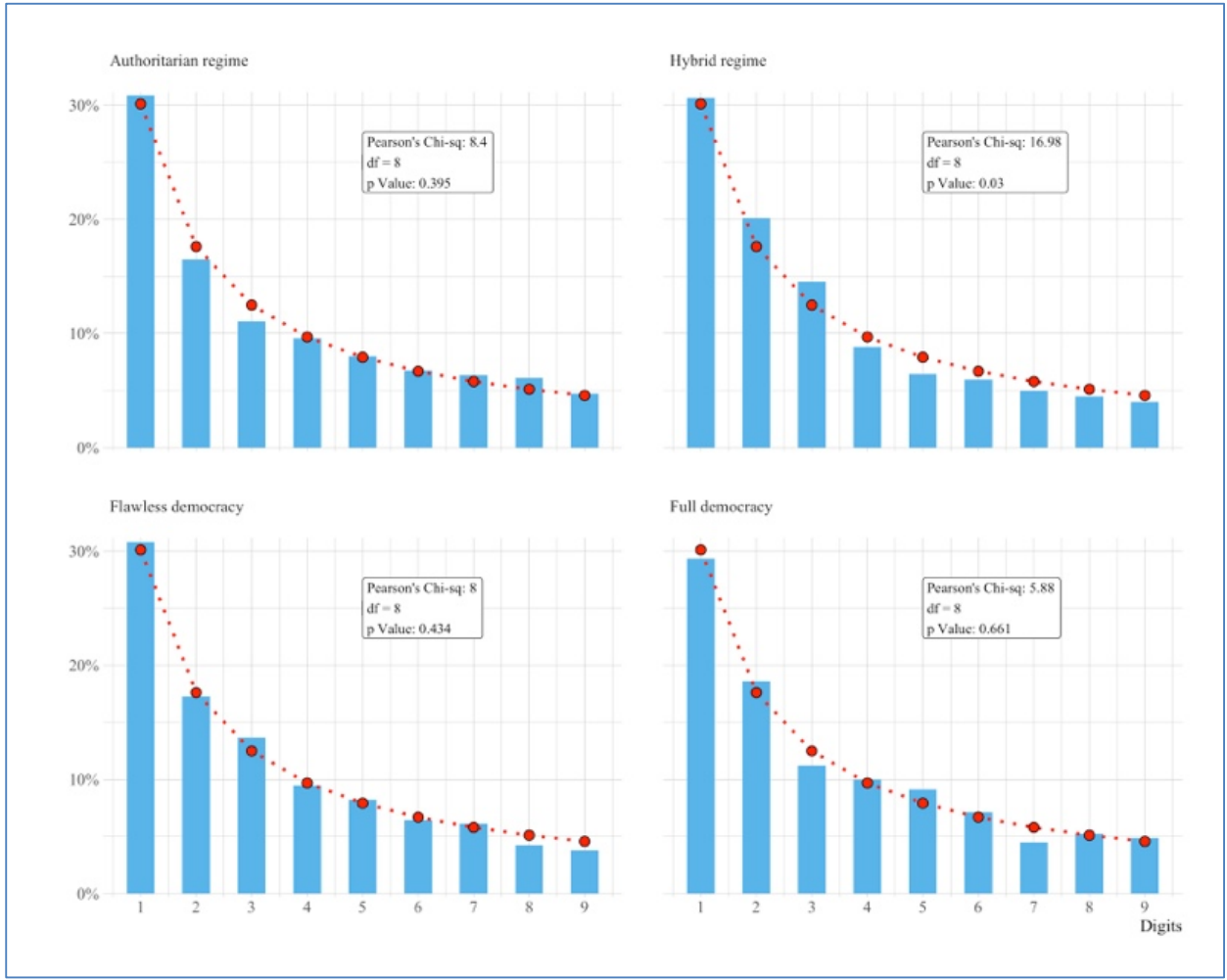
14

Total deaths

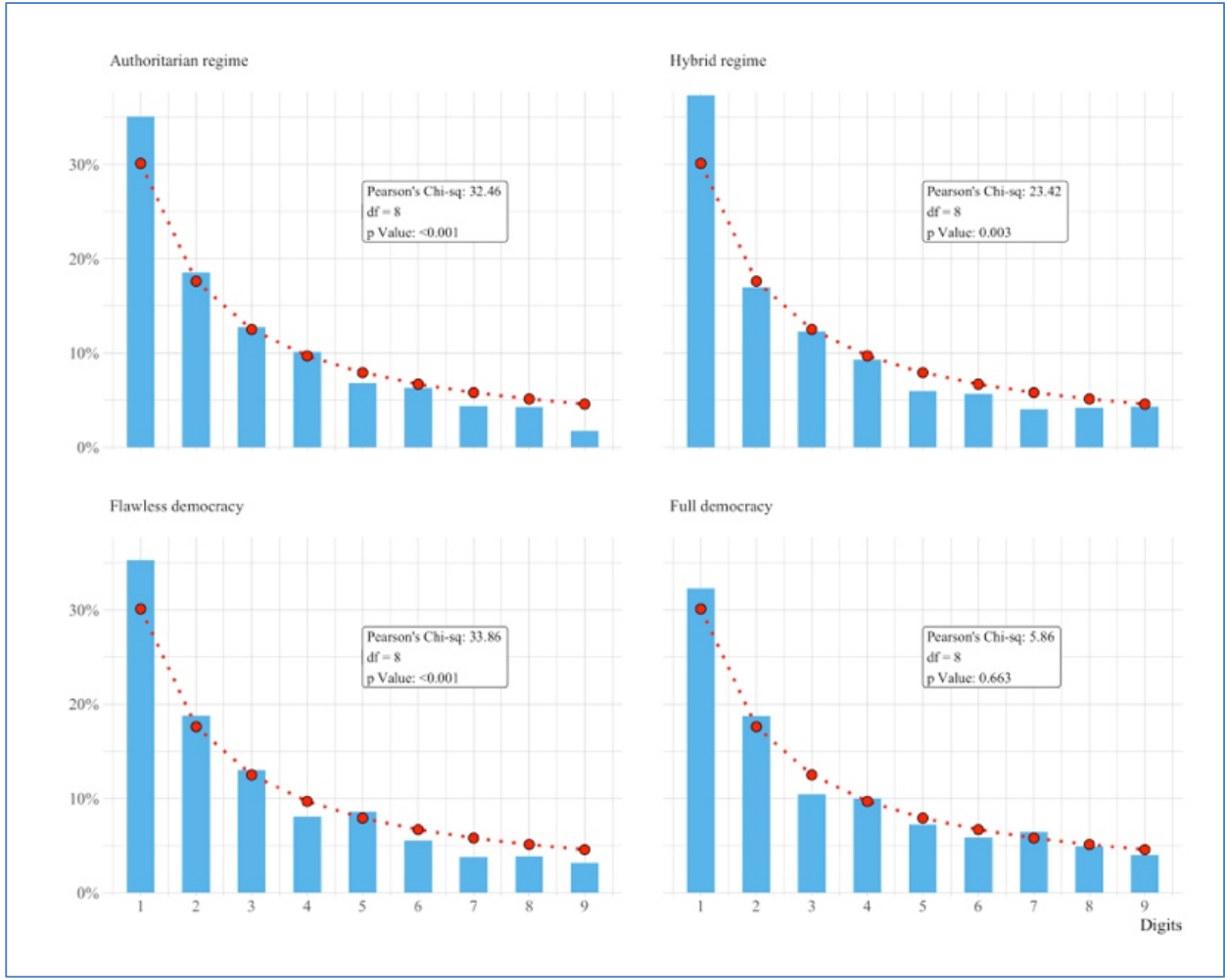

Notes: For each country, we only consider observations on total cases (deaths) on the date when the growth of the 7 day centered moving average ≥ 7.5% The blue bars are the observed frequency of the digits. The red dotted line is the expected frequency according the Benford's law. We compute the Pearson's Chi square and the p Value. For significance level of .01, the critical value is 20.09.

# References

1. Johns Hopkins University & Medicine. *COVID-19 Dashboard by the Center for Systems Science and Engineering (CSSE) at Johns Hopkins University (JHU)*, <https://coronavirus.jhu.edu/map.html> (2020).
2. The Economist. *Democracy Index 2019*, <https://www.eiu.com/topic/democracy-index> (2020).
3. Wadhams, N. & Jacobs, J. in *Bloomberg* (1 April 2020).
4. Wood, G. in *The Atlantic* (9 March 2020).
5. Firdaus, F. & Ratcliffe, R. in *The Guardian* (26 March 2020).
6. Rosenthal, E. in *New York Times* (25 May 2020).
7. Holz, C. A. China's Statistical System in Transition: Challenges, Data Problems, and Institutional Innovations. *Review of Income and Wealth* **50**, 381-409, doi:10.1111/j.0034-6586.2004.00131.x (2004).
8. Henderson, J. V., Storeygard, A. & Weil, D. N. Measuring Economic Growth from Outer Space. *American Economic Review* **102**, 994-1028, doi:10.1257/aer.102.2.994 (2012).
9. Magee, C. S. & Doces, J. A. Reconsidering regime type and growth: lies, dictatorships, and statistics. *International Studies Quarterly* **59**, 223-237 (2015).
10. Martinez, L. R. How Much Should We Trust the Dictator's GDP Growth Estimates? *Available at SSRN 3093296* (2019).
11. Merridale, C. The 1937 census and the limits of Stalinist rule. *The Historical Journal* **39**, 225-240 (1996).
12. Egorov, G., Guriev, S. & Sonin, K. Why resource-poor dictators allow freer media: A theory and evidence from panel data. *American political science Review* **103**, 645-668 (2009).
13. Gehlbach, S. & Sonin, K. Government control of the media. *Journal of Public Economics* **118**, 163-171 (2014).
14. Lorentzen, P. China's strategic censorship. *American Journal of Political Science* **58**, 402-414 (2014).
15. Guriev, S. & Treisman, D. Informational autocrats. *Journal of Economic Perspectives* **33**, 100-127 (2019).
16. Pogue, J. M., Devereaux, P. J., Thorlund, K. & Yusuf, S. Central statistical monitoring: Detecting fraud in clinical trials. *Clinical Trials* **10**, 225-235, doi:10.1177/1740774512469312 (2013).
17. Chilundo, B., Sundby, J. & Aanestad, M. Analysing the quality of routine malaria data in Mozambique. *Malaria Journal* **3**, 3, doi:10.1186/1475-2875-3-3 (2004).
18. Laxminarayan, R., Reif, J. & Malani, A. Incentives for reporting disease outbreaks. *PloS one* **9** (2014).
19. Buyse, M. *et al.* The role of biostatistics in the prevention, detection and treatment of fraud in clinical trials. *Statistics in medicine* **18**, 3435-3451 (1999).
20. Taylor, R. N., McEntegart, D. J. & Stillman, E. C. Statistical techniques to detect fraud and other data irregularities in clinical questionnaire data. *Drug information journal* **36**, 115-125 (2002).
21. Evans, S. Fraud and misconduct in medical science. *Encyclopaedia of Biostatistics, Wiley, Chichester*, 1583 (1998).
22. Al-Marzouki, S., Evans, S., Marshall, T. & Roberts, I. Are these data real? Statistical methods for the detection of data fabrication in clinical trials. *BMJ* **331**, 267, doi:10.1136/bmj.331.7511.267 (2005).

23  Elsa, V.-M., Jemma, H. C., Martin, L. & Jane, A. A key risk indicator approach to central statistical monitoring in multicentre clinical trials: method development in the context of an ongoing large-scale randomized trial. *Trials* **12**, A135, doi:10.1186/1745-6215-12-S1-A135 (2011).
24  Bailey, K. R. Detecting fabrication of data in a multicenter collaborative animal study. *Controlled Clinical Trials* **12**, 741-752, doi:https://doi.org/10.1016/0197-2456(91)90037-M (1991).
25  Collins, M. *et al.* Statistical techniques for the investigation of fraud in clinical research'. *Report of the ABPI Fraud Statistics Working Party* (1993).
26  Evans, S. 14: Statistical aspects of the detection of fraud. *Fraud and misconduct in biomedical research*, 186 (2001).
27  Benford, F. The law of anomalous numbers. *Proceedings of the American philosophical society*, 551-572 (1938).
28  Hill, T. P. A Statistical Derivation of the Significant-Digit Law. *Statistical Science* **10**, 354-363 (1995).
29  Knatterud, G. L. *et al.* Guidelines for Quality Assurance in Multicenter Trials: A Position Paper. *Controlled Clinical Trials* **19**, 477-493, doi:https://doi.org/10.1016/S0197-2456(98)00033-6 (1998).
30  Baigent, C., Harrell, F. E., Buyse, M., Emberson, J. R. & Altman, D. G. Ensuring trial validity by data quality assurance and diversification of monitoring methods. *Clinical Trials* **5**, 49-55, doi:10.1177/1740774507087554 (2008).
31  Venet, D. *et al.* A statistical approach to central monitoring of data quality in clinical trials. *Clinical Trials* **9**, 705-713, doi:10.1177/1740774512447898 (2012).
32  George, S. L. & Buyse, M. Data fraud in clinical trials. *Clinical investigation* **5**, 161 (2015).
33  Walker, M. E., Anonson, J. & Szafron, M. Economist intelligence unit democracy index in relation to health services accessibility: a regression analysis. *International Health* **7**, 49-59, doi:10.1093/inthealth/ihu064 (2014).
34  Lim, C. & MacGregor, C. A. Religion and Volunteering in Context: Disentangling the Contextual Effects of Religion on Voluntary Behavior. *American Sociological Review* **77**, 747-779, doi:10.1177/0003122412457875 (2012).
35  Elkomy, S., Ingham, H. & Read, R. Economic and political determinants of the effects of FDI on growth in transition and developing countries. *Thunderbird International Business Review* **58**, 347-362 (2016).
36  Freedom House. *Freedom in the World*, <https://freedomhouse.org/report/freedom-world> (2020).
37  Varieties of Democracy. *V-Dem Dataset - Version 10*, <https://www.v-dem.net/en/data/data-version-10/> (2020).
38  Center for Systemic Peace, P. P. *Polity5: Political Regime Characteristics and Transitions, 1800-2018*, <http://www.systemicpeace.org/inscrdata.html> (2020).
39  Ritchie, H., Orrtiz-Ospina, E., Roser, M. & Hasell, J. *COVID-19 deaths and cases: how do sources compare?*, <https://ourworldindata.org/covid-sources-comparison> (19 March 2020).
40  The World Bank. *World Bank Open Data*, <https://data.worldbank.org/> (2020).
41  Zitzewitz, E. Forensic Economics. *Journal of Economic Literature* **50**, 731-769, doi:10.1257/jel.50.3.731 (2012).
42  Varian, H. R. Benfords law. *American Statistician* **26**, 65-& (1972).
43  Nigrini, M. & Wood, W. Assessing the integrity of tabulated demographic data. 1995. *Preprint* (1995).
44  Nigrini, M. J. A taxpayer compliance application of Benford's law. *The Journal of the American Taxation Association* **18**, 72 (1996).

45  Ley, E. On the peculiar distribution of the US stock indexes' digits. *The American Statistician* **50**, 311-313 (1996).
46  Durtschi, C., Hillison, W. & Pacini, C. The effective use of Benford's law to assist in detecting fraud in accounting data. *Journal of forensic accounting* **5**, 17-34 (2004).
47  Tam Cho, W. K. & Gaines, B. J. Breaking the (Benford) Law. *The American Statistician* **61**, 218-223, doi:10.1198/000313007X223496 (2007).
48  Diekmann, A. Not the First Digit! Using Benford's Law to Detect Fraudulent Scientif ic Data. *Journal of Applied Statistics* **34**, 321-329, doi:10.1080/02664760601004940 (2007).

# Authoritarian Governments Appear to Manipulate COVID Data

## Supplement

Mudit Kapoor, Anup Malani, Arnav Agarwal, Shamika Ravi

**Further evidence on excess burden in democracies**

### Figure S1

Total COVID cases, deaths per million people, and Case fatality ratio, by government regime, over time.

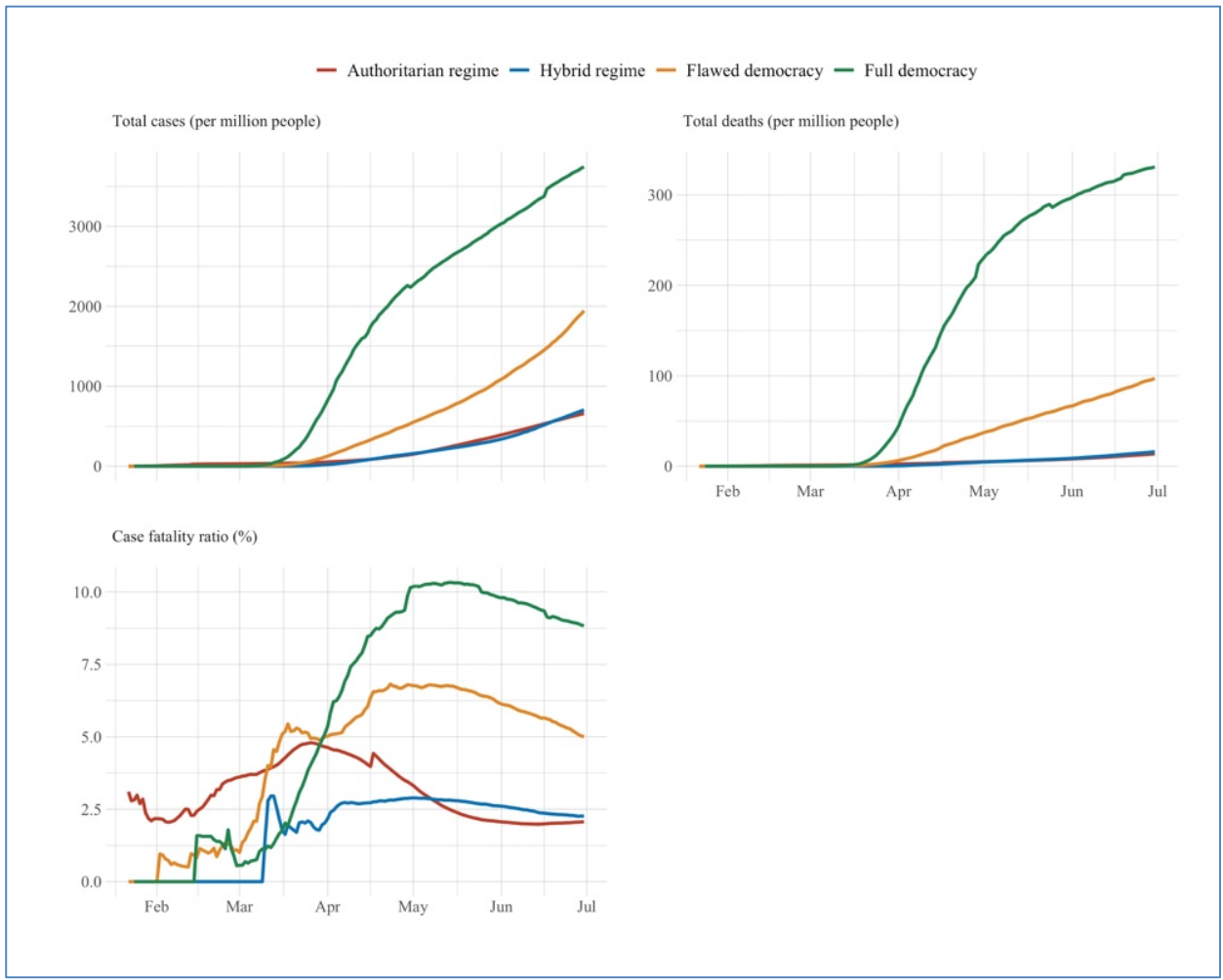

**Further evidence on muted variation around moving average (test 1)**

**Figure S2**

New daily cases and deaths per million people and 7-day moving average of the same, by government regime.

Daily cases

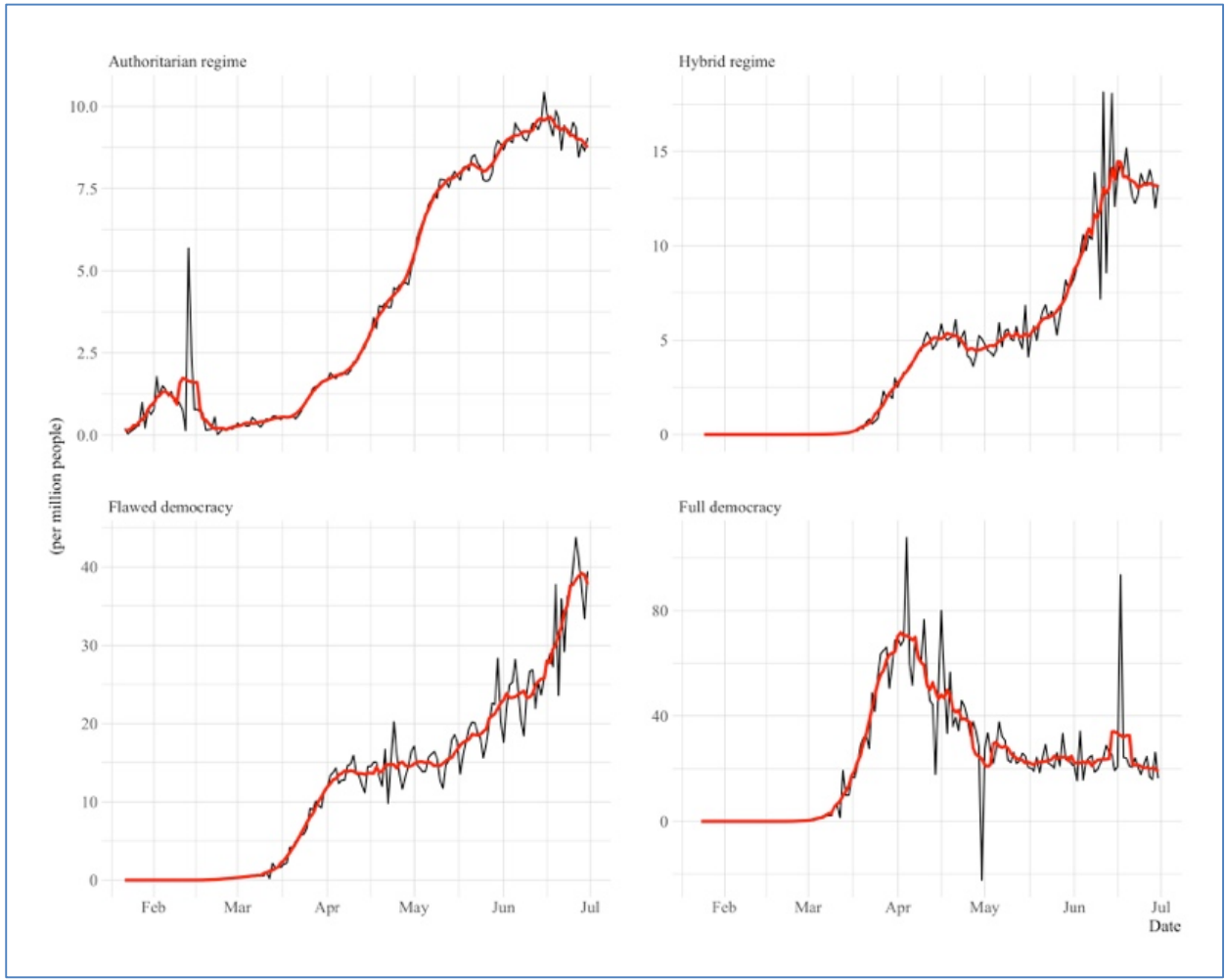

Daily deaths

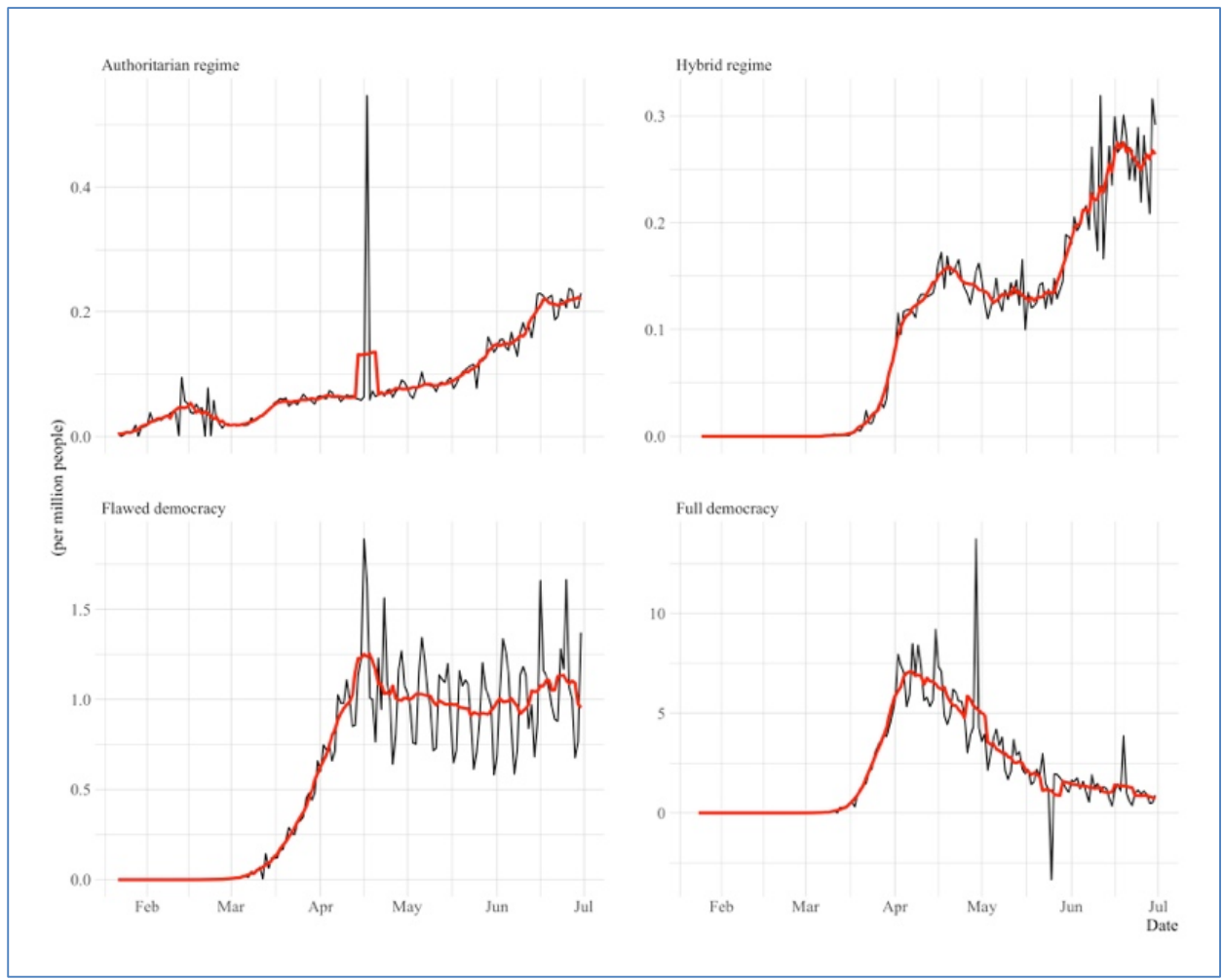

Notes. Case and deaths data are from Johns Hopkins University. New cases (deaths) are calculated by summing new cases (deaths) by day across all countries with a given regime type. Moving average is the 7 day centered moving average.

## Table S1

Ordinary least squares regression of deviations from a moving average on measures of democracy.

| Daily cases | | | | |
|---|---|---|---|---|
| | EIU democracy score | FH democracy score | VD democracy score | Polity5 |
| Democracy score | -0.01 | 0.04 | -0.00 | 0.08 |
| | (-0.22 - 0.20) | (-0.12 - 0.20) | (-0.19 - 0.18) | (-0.06 - 0.21) |
| Per capita GDP | 0.82*** | 0.81*** | 0.82*** | 0.88*** |
| | (0.54 - 1.11) | (0.53 - 1.09) | (0.54 - 1.11) | (0.58 - 1.19) |
| Trade as % of GDP | -0.27 | -0.29 | -0.29 | -0.32 |
| | (-1.01 - 0.48) | (-1.05 - 0.48) | (-1.03 - 0.45) | (-1.08 - 0.43) |
| Health as % of GDP | 0.87** | 0.81** | 0.86** | 0.74* |
| | (0.09 - 1.65) | (0.01 - 1.60) | (0.06 - 1.67) | (-0.05 - 1.54) |
| Share of ≥65 years in pop | -0.87** | -0.96** | -0.87** | -1.13*** |
| | (-1.61 - -0.13) | (-1.70 - -0.22) | (-1.58 - -0.16) | (-1.91 - -0.35) |
| Country ≥40° N latitude | 0.85* | 0.89* | 0.87* | 0.92* |
| | (-0.14 - 1.83) | (-0.09 - 1.86) | (-0.11 - 1.85) | (-0.12 - 1.95) |
| Constant | -4.32** | -4.08** | -4.25** | -4.49** |
| | (-7.74 - -0.89) | (-7.60 - -0.56) | (-7.78 - -0.72) | (-8.03 - -0.95) |
| N | 19,430 | 19,223 | 19,230 | 18,740 |
| R2 | 0.10 | 0.10 | 0.10 | 0.10 |

| Daily deaths | | | | |
|---|---|---|---|---|
| | EIU democracy score | FH democracy score | VD democracy score | Polity5 |
| Democracy score | 0.16** | 0.15*** | 0.17*** | 0.18*** |
| | (0.03 - 0.29) | (0.06 - 0.25) | (0.06 - 0.28) | (0.07 - 0.29) |
| Per capita GDP | 0.30*** | 0.32*** | 0.29*** | 0.44*** |
| | (0.13 - 0.47) | (0.15 - 0.48) | (0.13 - 0.46) | (0.26 - 0.61) |
| Trade as % of GDP | 0.57** | 0.55** | 0.56** | 0.52** |
| | (0.08 - 1.05) | (0.06 - 1.04) | (0.07 - 1.04) | (0.05 - 0.99) |
| Health as % of GDP | 0.98*** | 0.86*** | 0.83*** | 0.85*** |
| | (0.50 - 1.46) | (0.39 - 1.34) | (0.36 - 1.30) | (0.38 - 1.31) |
| Share of ≥65 years in pop | -0.58** | -0.61*** | -0.56*** | -0.70*** |
| | (-1.03 - -0.12) | (-1.05 - -0.18) | (-0.97 - -0.15) | (-1.15 - -0.26) |
| Country ≥40° N latitude | 0.49 | 0.47 | 0.47 | 0.45 |
| | (-0.13 - 1.10) | (-0.13 - 1.07) | (-0.15 - 1.08) | (-0.16 - 1.05) |
| Constant | -8.24*** | -8.01*** | -7.69*** | -8.93*** |
| | (-10.33 - -6.15) | (-10.11 - -5.91) | (-9.78 - -5.61) | (-11.04 - -6.83) |
| N | 19,430 | 19,223 | 19,230 | 18,740 |
| R2 | 0.17 | 0.18 | 0.18 | 0.18 |

Note: *** p<0.01, ** p<0.05, * p<0.1. 95% Confidence intervals are in parenthesis. The errors are clustered at the country level. The dependent variable is the natural logarithm of the squared deviation of the observed value from the 7 day centered moving average plus one per million people for each country on a daily basis, from the date when the first case was noted till June 30, 2020. Freedom House democracy score ranges from 0 to 100, to make it comparable to the EIU democracy score, the score is divided by 10. The VDEM score ranges from 0 to 1, to make it comparable to EIU democracy score, it is multiplied by 10. Similarly the modified polity5 score ranges from -10 (strongly autocratic) to +10 (strongly democratic), therefore, to make it comparable we add 10 to the score and divide it by 2.

# Further evidence on Benford's Law (Test 2)
## Table S2

| Total cases | | | | |
|---|---|---|---|---|
| | Growth rate = 5% | | Growth rate = 10% | |
| | Pearson Chi-square | P Value | Pearson Chi-square | p Value |
| Authoritarian regime | 6.98 | 0.54 | 9.10 | 0.33 |
| Hybrid regime | 9.31 | 0.32 | 15.84 | 0.04 |
| Flawed democracy | 9.36 | 0.31 | 6.84 | 0.55 |
| Full democracy | 6.60 | 0.58 | 5.15 | 0.74 |

| Total deaths | | | | |
|---|---|---|---|---|
| | Growth rate = 5% | | Growth rate = 10% | |
| | Pearson Chi-square | P Value | Pearson Chi-square | p Value |
| Authoritarian regime | 15.77 | 0.05 | 49.50 | p < .01 |
| Hybrid regime | 14.89 | 0.06 | 23.71 | p < .01 |
| Flawed democracy | 29.10 | 0.00 | 37.03 | p < .01 |
| Full democracy | 5.79 | 0.67 | 6.24 | 0.62 |